\documentstyle[12pt]{article}

\begin{document}

\hfill{UTTG-16-94}
\baselineskip=24pt

\begin{center}
{\large{\bf Strong Interactions at Low Energies}}\\

\vspace{36pt}

Steven Weinberg\footnote{Research supported in part by the
Robert A. Welch
 Foundation and NSF Grant PHY 9009850.  E-mail address:
weinberg@physics.utexas.edu.}

\vspace{4pt}
Theory Group, Department of Physics, University of Texas\\
Austin, TX, 78712\\

\end{center}

Effective field theories are playing an increasing role in
the study of a wide variety of physical phenomena, from $W$
and $Z$ interactions to superconductivity.  Regarding the
subject of this talk, we have known for years that the low
energy strong interactions of nucleons and pions are well
described in the tree approximation\footnote{In dealing with
nuclear forces, the tree approximation must be applied to
the nucleon-nucleon potential, rather than to the scattering
amplitude.} by an effective field theory, with Lagrangian[1]
\begin{eqnarray}
&& {\cal L}_{\rm eff}=-\frac{\partial_\mu\vec{\pi}\cdot
\partial^\mu \vec{\pi}}{2(1+\vec{\pi}^2/F_\pi^2)^2} -
\frac{m_\pi^2\,\vec{\pi}^2}{2(1+\vec{\pi}^2/F_\pi^2)^2}
\nonumber\\&&+\bar{N}\left[i\partial_0-\frac{2
\vec{t}\cdot(\vec{\pi}\times \partial_0 \vec{\pi})}{F_\pi^2
(1+\vec{\pi}^2/F_\pi^2)^2} -m_N-\frac{2g_A \vec{t}\cdot
(\mbox{\boldmath $\sigma\cdot
\nabla$})\vec{\pi}}{F_\pi(1+\vec{\pi}^2/F_\pi^2)}\right]N
\nonumber\\&&
-\frac{1}{2} C_S (\bar{N}N)^2-\frac{1}{2} C_T (\bar{N} \mbox
{\boldmath$\sigma$} N )^2
\end{eqnarray}
where $g_A=1.25$ and $F_\pi=190\;{\rm MeV}$, and $C_S$ and
$C_T$ are constants whose values can be fit to the two
nucleon-nucleon scattering lengths.  (Spatial vectors
are boldface; arrows denote isovectors.)  In this talk I
will  describe some current research that takes us beyond
the leading terms provided by Eq. (1), in three different
directions.

\vspace{24pt}
\noindent
{\bf 1 Isospin Breaking Corrections}\\
Using Eq. (1) in the tree approximation gives just the first
term in an expansion in powers of $q$, the typical value of
the pion and nucleon three-momenta and pion mass.  Higher
terms in the expansion are generated[2]  by including
more derivatives in ${\cal L}_{\rm eff}$, each of which
contributes a factor of order $q$, or more nucleon fields,
each of which contributes a factor of order $q^{1/2}$, or
more factors of $u$ and $d$ quark masses, each of
which contributes a factor of order $q^2$ (because
$m_\pi^2\propto m_u+m_d$), or loops in the Feynman graphs,
each of which contributes a factor $q^2$.  These corrections
have been explored in great detail (including also  strange
particles), especially by Gasser and Leutwyler[3] for pions
and single nucleons, and more recently by Bernard, Kaiser,
and Meissner[4] for pion photoproduction and by Ordo\~{n}ez
and van Kolck[5] for multinucleon problems.  Here I want to
concentrate on the quark mass corrections, which produce
violations of isospin conservation.  This is of renewed
interest now, because as I learned from Aron Bernstein there
are plans to measure the $\pi^0$-nucleon scattering length,
which is sensitive to isospin violating terms in the
effective Lagrangian.

The quark mass terms in quantum chromodynamics may be put in
the form
\begin{equation}
{\cal L}_{\rm mass}=-(m_u+m_d)V_4 - (m_u-m_d)A_3
\end{equation}
where
\begin{equation}
V_4=\frac{1}{2}(\bar{u}u+\bar{d}d)~~~~~~~~A_3=\frac{1}{2}
(\bar{u}u-\bar{d}d)\;.
\end{equation}
The operators $V_4$ and $A_3$ are spatial scalars, and
components of independent chiral four-vectors $A_\alpha$ and
$V_\alpha$.  We must add terms to the effective Lagrangian
with these transformation properties.  From just the pion
field alone (with no derivatives) we can construct no term
$A_3$ and just one term $V_4$, the term in (1) proportional
to $m_\pi^2$.
{}From pion fields and a nucleon bilinear (but no derivatives)
we can construct only one term of each type
$$V_4\propto \left(\frac{1-
\vec{\pi}^2/F_\pi^2}{1+\vec{\pi}^2/F_\pi^2}\right)\bar{N}N
$$
$$
A_3\propto \bar{N}t_3 N
-
\frac{2}{F_\pi^2}\left(\frac{\pi_3}{1+\vec{\pi}^2/F_\pi^2}
\right)\bar{N}\vec{t}\cdot\vec{\pi} N
$$
where $N$ is the nucleon doublet.  Therefore in the
effective Lagrangian
we must include a term
\begin{eqnarray}
&&\delta{\cal L}_{eff}=-\frac{A}{2}\left(\frac{1-
\vec{\pi}^2/F_\pi^2}{1+\vec{\pi}^2/F_\pi^2}\right)\bar{N}N
\nonumber\\
&&-B\left[\bar{N}t_3 N-
\frac{2}{F_\pi^2}\left(\frac{\pi_3}{1+\vec{\pi}^2/F_\pi^2}
\right)\bar{N}\vec{t}\cdot\vec{\pi} N\right]
\end{eqnarray}
where $A$ and $B$ are constants proportional to the
coefficients in (2):
\begin{equation}
A\propto m_u+m_d~~~~~~~~~~~~B\propto m_u-m_d\;.
\end{equation}
The pion-nucleon terms in (1) have single derivatives, which
contribute factors of order $q$, while the quark masses in
(4) contribute factors of order $q^2$, so the effects of (4)
are leading corrections, suppressed by just {\em one} factor
of $q$.

The terms in (4) make a contribution to the scattering
length
for the pion-nucleon scattering process
$\pi_a+N\rightarrow\pi_b+N$ (written as a matrix in the
isospin space of the nucleon):
\begin{equation}
\delta a_{ba}=\frac{1}{4\pi[1+m_\pi/m_N]}\;\times\;\left[
\frac{2A}{F_\pi^2}\delta_{ab}
\;+\;\frac{2B}{F_\pi^2}(t_a\delta_{3b}+t_b\delta_{3a})\right
]\;.
\end{equation}
The $A$ term is the notorious $\sigma$-term.  The $B$ term
was also described years ago[6].  What (I think) is new here
is the full isospin-breaking term[7] in the effective action
(4), which allows us easily to calculate the effect of
isospin violations in other processes, such as
$\pi+N\rightarrow\pi+\pi+N$.  One immediate consequence of
(4) is that isospin violation never appears in any process
that does not involve at least one neutral pion.

Inspection of (4) shows that the constants $A$ and $B$ are
related to the shifts $\delta m_n$ and $\delta m_p$ in the
nucleon masses due to the quark masses:
\begin{eqnarray}
A&=&\delta m_p+\delta m_n=(m_u+m_d)\;<
p|(\bar{u}u+\bar{d}d|p>
\nonumber\\B&=&\delta m_p-\delta m_n=(m_u-m_d)\;<
p|(\bar{u}u-\bar{d}d|p>\;.
\end{eqnarray}
Unfortunately the nucleon expectation value of
$\bar{u}u+\bar{d}d$ is not related in any simple way to
observable quantities, so it is not possible to calculate
$A$ without dynamical assumptions.  On the other hand, $B$
is given by $SU(3)$ symmetry as:
\begin{equation}
B\simeq \left(\frac{m_u-m_d}{m_s}\right)(m_\Xi-
m_\Sigma)\approx -2.5\;{\rm MeV}\;.
\end{equation}
This satisfies an important consistency condition.  The full
proton-neutron mass difference is the sum of $B$ and an
electromagnetic term, which is almost certainly positive, so
we must have $B<m_p-m_n=-1.3$ MeV, and we do.  It will be
very interesting to see if experiments on low energy
$\pi^0$-nucleon interactions confirm these predictions.

\vspace{24pt}
\noindent
{\bf 2 General Effective Lagrangians}\\
The structure of the effective Lagrangian (1) is dictated by
its invariance under $SU(2)\times SU(2)$ spontaneously
broken to $SU(2)$, which induces on the pion field the non-
linear symmetry transformation[8]:
\begin{equation}
\delta \vec{\pi}=\vec{\epsilon}(1- \vec{\pi}^2)+2
\vec{\pi}(\epsilon\cdot \vec{\pi})\;.
\end{equation}
This was generalized by Callan,  Coleman, Wess, and
Zumino[9] to any group $G$ broken to any subgroup $H\subset
G$, in which case an element $g\in G$ induces on the general
Goldstone boson fields $\pi^a$ the transformation
$\pi\rightarrow\pi'$, defined by
\begin{equation}
gU(\pi)=U(\pi')h(\pi,g)\;,
\end{equation}
where $h\in H$ and $U(\pi)$ is a representative of the coset
space $G/H$, parameterized by the Goldstone boson fields.
Now, we know how to construct $G$-invariant Lagrangian
densities our of covariant derivatives of $\pi^a$, but this
is not the most general possibility.  We also can have a
Lagrangian density that under $G$ transformations changes by
a derivative
$$
{\cal L}\rightarrow {\cal L}+\partial_\mu {\cal F}^\mu
$$
so that the action is still invariant.  Wess and Zumino[10]
pointed out that the ABJ anomaly from fermion loops yields
such a term in the effective Lagrangian for the case of
$SU(3)\times SU(3)$ spontaneously broken to $SU(3)$:
\begin{equation}
{\cal L}_{\rm
WZ}=\frac{1}{6\pi^2}\epsilon^{\mu\nu\rho\sigma}\,{\rm
Tr}\,\{\Pi\,\partial_\mu\Pi\,\partial_\nu\Pi\,\partial_\rho
\Pi\,\partial_\sigma\Pi\,\}\;+\;O(\Pi^6)
\end{equation}
where
$$
\Pi\equiv\frac{1}{2}\lambda_a\pi^a\;,
$$ $\lambda_a$ are the Gell-Mann matrices (with ${\rm
Tr}\,\lambda_a^2=2$), and the coset representatives $U(\pi)$
are chosen as
\begin{equation}
U(\pi)=e^{i\Pi}\;.
\end{equation}
Witten[11] then showed that although the correction $\int
d^4x {\cal L}_{\rm WZ}$ to the action is not the integral of
a $G$-invariant Lagrangian density over spacetime,  it {\em
is} the integral of a $G$-invariant Lagrangian density
${\cal L}_{\rm WZW}$ over a five-dimensional ball that has
four-dimensional spacetime (Euclideanized and compactified
to a four-sphere) as its boundary.  This raises the question
whether there are any other terms in the effective
Lagrangian density, not necessarily related to ABJ
anomalies, that although not invariant under $G$
nevertheless yield $G$-invariant contributions to the
action.  Have we been missing something?

This question has now been answered by Eric D'Hoker and
myself[12], with help at the start from Eddie Farhi.  Our
analysis is in four steps:

\vspace{6pt}
\noindent
({\bf a}) As in ref. [11], we first compactify spacetime to
a sphere $S_4$ by assuming that all fields approach definite
limits for $x^\mu\rightarrow
\infty$.  If the
homotopy group $\pi_4(G/H)$ is trivial (as is the case for
$SU(N)\otimes SU(N)$ spontaneously broken to $SU(N)$), or if
$U(\pi(x))$ belongs to the trivial element
of $\pi_4(G/H)$, then we may introduce a smooth function
$\tilde{\pi}^a(x,t_1)$, such that
$$\tilde{\pi}^a(x,1)=\pi^a(x)~~~~~~~\tilde{\pi}^a(x,0)=0.$$
In this way spacetime is extended to a five-ball $B_5$ with
boundary $S_4$ and coordinates $x^\mu$ and $t_1$.
The
action may then be written in the five-dimensional form
$$
S[\pi]=\int_{B_5}d^4x\, dt_1\;{\cal L}_1
$$
where
$$
 {\cal L}_1\equiv\frac{\delta I[\tilde{\pi}]}{\delta
\tilde{\pi}^a(x,t_1)}\;\frac{\partial\tilde{\pi}^a(x,t_1)}
{\partial t_1}\;.
$$

\vspace{6pt}
\noindent
({\bf b}) It is straightforward to show that if the action
$I$ is invariant under $G$, then the density ${\cal L}_1$ is
also invariant under $G$.  Thus any $G$-invariant term in
the action can be written in the Witten form, as a
five-dimensonal integral of an invariant density.

\vspace{6pt}
\noindent
({\bf c}) From the definition of ${\cal L}_1$ in terms of
$\delta I/\delta \pi$,  we learn not only that it is $G$-
invariant, but also that it satisfies an integrability
condition, which implies that ${\cal L}_1$ is a component of
a $G$-invariant closed five-form $\Omega_5$.  That is, in
the language of differential forms:
\begin{equation}
I=\int_{B_5} \Omega_5   ~~~~~~~~~~ d\Omega_5=0\;.
\end{equation}
Now, if $\Omega_5$ is exact, then the four-form $F_4$
satisfying $\Omega_5=dF_4$ can be chosen to be
$G$-invariant, in which case $I$ is the four-dimensional
integral $\int_{S_4}F_4$ over $S_4$ of a $G$-invariant
Lagrangian density.  Hence the allowed terms in the four-
dimensional Lagrangian density that are {\em not} $G$-
invariant are in one-to-one correspondence with closed
five-forms, modulo exact five-forms.
These are the generators of the fifth de Rham cohomology
$H^5(G/H;{\bf R})$ of the space $G/H$.

\vspace{6pt}
\noindent
({\bf d}) It only remains to find the five-forms that
generate
$H^5(G/H;{\bf R})$.  These are all known where $G/H$ is
itself a simple group.  If  $G/H=SU(N)$ with $N\geq 3$
then $H^5(G/H;{\bf R})$ has a single generator:
\begin{eqnarray}
&&\Omega_5=\frac{i}{240\pi^2} {\rm Tr}\,\left\{ U^{-
1}\,dU\;\wedge\;  U^{-1}\,dU\;\wedge\; U^{-
1}dU\right.\nonumber\\&&~~~~~~~~\left.\;\wedge\; U^{-
1}\,dU\;\wedge\; U^{-1}\,dU\right\}\;.
\end{eqnarray}
For the QCD case of $SU(3)\times SU(3)$ spontaneously broken
to $SU(3)$, we have
$G/H=SU(3)$, and the unique generator (15) is the
Wess-Zumino-Witten five dimensional Lagrangian density.  So
at
least as far as the strong interactions at low energy are
concerned, we have not been missing anything.
Where $G/H$ is any simple Lie group
other than $SU(N)$ with $N\geq 3$, the cohomology is
trivial, and so the four-dimensional Lagrangian density must
be $G$-invariant.  This includes the original case of
$SU(2)\times SU(2)$ spontaneously broken to $SU(2)$, where
$G/H=SU(2)$.

Of course, a great deal is known about the fifth cohomology
groups
$H^5(G/H;{\bf R})$ even where $G/H$ is not a simple Lie
group.
One interesting result is that if $G$ itself is one of the
simple Lie groups other than $SU(N)$ with $N\geq 3$, then
$G/H$ has trivial fifth cohomology group for any subgroup
$H\subset G$, and so the Lagrangian density must be
$G$-invariant.
These groups $G$ also have vanishing triangle anomalies
for fermions in any representation of $G$, because in all
representations the generators $t_\alpha$ satisfy
$${\rm Tr}\,\{t_\alpha(t_\beta t_\gamma+t_\gamma
t_\beta)\}=0$$
so here we would not have expected a Wess-Zumino-Witten term
anyway.  D'Hoker is continuing with the study of general
coset spaces $G/H$, to map out the detailed relation between
the possibility of ABJ anomalies and non-$G$-invariant terms
in the four-dimensional Lagrangian density.  It is
remarkable that in all the cases we have studied, the
possibility of anomalies could have been discovered (or
ruled out) within the effective field theory of soft
Goldstone bosons, without ever looking at a fermion loop.

\vspace{24pt}
\noindent
{\bf 3 The Nonrelativistic Quark Model: Sum Rules {\it vs}
Large $N_c$}  \\
The derivation of results of the non-relativistic quark
model from quantum chromodynamics has long remained
problematical.  Recently, the large $N_c$ approximation[13]
has been used[14-18] to derive some of the quark model
results for baryons.  Specifically, it is found that the
nucleon doublet is
connected by one-pion transitions to a  `tower' of narrow
baryon
states, with spin and isospin $J=T=1/2, 3/2, \cdots N_c/2$,
and positive parity.  According to this picture, the
amplitudes for pion transitions between the tower states
together with the spin and isospin operators form a
contracted $SU(4)$ algebra[15-17], under which the baryon
tower transforms irreducibly.  The one-pion transition
amplitudes derived in this way are just those of the non-
relativistic quark model.

I want to point out that strikingly similar results can be
derived in a very different way.  In this approach no use is
made of the large $N_c$ approximation or dynamical models
like the non-relativistic quark model or Skyrme model,
beyond the qualitative assumption that there is a tower of
narrow baryon
states, with spin and isospin $J=T=1/2, 3/2, \cdots N_c/2$,
and positive parity, that are connected only to each other
by one-pion transitions.  By the use of well-known sum rules
saturated with narrow tower states, we will be able to show
(1) that the tower states are degenerate\footnote{This is a
somewhat surprising result.  In general for large $N_c$ it
is only the lower tower states with $T=J=O(1)$ that become
degenerate when $N_c\rightarrow\infty$.  The work described
here shows that in order to understand the splittings of the
baryon tower masses, it will be necessary to take into
account single-pion transitions from tower to non-tower
states, rather than $1/N_c$ corrections to matrix elements
between tower states.},  and (2)  that
the pion transition amplitudes are part of an {\em
uncontracted} $SU(4)\times O(3)$
Lie algebra, under which the baryonic tower transforms as a
symmetric rank-$N_c$ tensor\footnote{Of
course, if one also assumes
that  $N_c$  is large, the leading terms in the matrix
elements of the $SU(4)$ generators between low tower states
in this representation
will grow as $N_c$.  Since the spin and isospin matrices are
for these states are of order $N_c^0$, they and the  leading
terms in the pion
transition amplitudes for $N_c\rightarrow \infty$ will
furnish a contracted $SU(4)$ algebra, as found in references
[15] - [17].} of $SU(4)$ and a  singlet under $SO(3)$.  Here
$N_c$ is
any integer, including $N_c=3$; we keep $N_c$ a free
parameter to facilitate comparison with work based on the
large $N_c$ limit.  By relying hardly at all here on the
large $N_c$ approximation, this approach offers a prospect
of a more convincing derivation of the main results of the
non-relativistic quark model.

To derive these results, lets first recall the relevant sum
rules and their algebraic consequences.   We usually think
of spontaneously broken symmetries like
$SU(2)\times SU(2)$ as being manifested entirely in low
energy theorems
for the interaction of Goldstone bosons.  There are similar
low-energy theorems for the interactions of soft photons.
But these low energy theorems when married to dispersion
relations yield sum rules, like the celebrated
Adler-Weisberger, Drell-Hearn, and Cabibbo-Radicati  sum
rules.  Other sum rules known as superconvergence relations
are provided by assumptions that limit the asymptotic
behaviour of scattering amplitudes at high energy.  When we
assume that these sum rules are saturated by any number of
particles and narrow
resonances, and write down all the sum rules for scattering
of Goldstone bosons and/or photons not only on stable
targets but also on all the resonances, we find that they
take remarkably similar algebraic forms[19-21].
This is a very old story, going back more years than I care
to remember.   The new thing I want to discuss here is the
solution of these sum rules under a specific assumption
about the menu of baryonic spins and isospins for general
$N_c$.

Consider pion scattering on an arbitrary hadronic target.
The saturated sum rules can be expressed in terms of the
matrix elements for pion transitions $\alpha\rightarrow
\beta+\pi_a$
between stable or resonant states $\alpha$, $\beta$ with
helicities
$\lambda$ and $\lambda'$.  For any such transition, we can
adopt a Lorentz frame in which the initial and final states
have {\it collinear} momenta
${\bf p}$ and ${\bf p}'$, say, in the
3-direction.    Using
invariance under rotations around the $3$-axis and boosts
along the $3$-axis, these matrix
elements may be written as:
\begin{eqnarray}
&&<\beta,{\bf p}^\prime, \lambda^\prime ;  \pi_a,q|{  }S{ }
|\alpha,{\bf p},\lambda>{ } \equiv \frac{ 2  (m_\alpha^2 -
m_\beta^2)}{(2\pi)^{9/2}(8q^0p^0p^{\prime
0})^{1/2}F_\pi}\nonumber\\&&
{}~~~~~~~~~~~~\times\;[X_a(\lambda)]_{\beta\alpha}\,\delta_{
\lambda'\lambda}
\delta^4(p^\prime+q-p)
\end{eqnarray}
with a coefficient
$[X_a(\lambda)]_{\beta\alpha}$ that is independent of $|{\bf
p}|$ and
$|{\bf p}'|$.  (The axial coupling $g_A\simeq
1.25$ is just the helicity $+1/2$ proton-neutron element of
the matrix $X_1+iX_2$.)  Parity conservation
tells us that
\begin{equation}
[X_a(-\lambda)]_{\beta\alpha}=-\Pi_\alpha\Pi_\beta(-
1)^{J_\alpha-
J_\beta}[X_a(\lambda)]_{\beta\alpha}
\end{equation}
where $\Pi_\alpha$ and $J_\alpha$ are the parity and spin of
the
baryonic state $\alpha$.
  Isospin invariance tells us that
$$[T_a,X_b(\lambda)]=i\epsilon_{abc}X_c(\lambda)~~{\rm
where}~~[T_a,T_b]=i\epsilon_{abc} T_c\;.$$
In this language, when all of the
Adler-Weisberger sum rules for scattering of a pion on all
single-hadron states (either stable particles or narrow
resonances) are saturated with single-hadron states, these
sum rules read  simply[19]
\begin{equation}
[X_a(\lambda),X_b(\lambda)]=i\epsilon_{abc}\,T_c\;.
\end{equation}
Thus for each helicity the reduced pion matrix elements
$X_a(\lambda)$ and the isospin matrix $T_a$ together form an
$SU(2)\times SU(2)$ algebra\footnote{This result can be
generalized to arbitrary groups $G$ broken to arbitrary
subgroups $H$: the reduced amplitudes for Goldstone boson
emission together with the unbroken symmetry generators
furnish a representation of the algebra of $G$.  This holds
even  where $G/H$ is not a symmetric space, i. e., where
terms linear in the broken as well as the unbroken
generators appear in the commutators of the broken
generators with each other.}.  There are also two
superconvergence relations that follow from the absence of
$T=2$ Regge
trajectories with $\alpha(0)>0$ in the cross channel.
One takes the form[19]
\begin{equation}
[X_a(\lambda),[X_b(\lambda),m^2]\propto \delta_{ab}\;.
\end{equation}
The other is a spin-flip superconvergence relation, that
connects different helicities[20]
\begin{eqnarray}
&&X_a(\lambda\pm 1)X_b(\lambda\pm 1)\,m-X_a(\lambda\pm
1)\,m\,X_b(\lambda)-X_b(\lambda\pm
1)\,m\,X_a(\lambda)\nonumber\\&&
{}~~~~~~~~~~~+m\,X_b(\lambda)X_a(\lambda)~~\propto~~
\delta_{ab}
\end{eqnarray}
where $m$ is the hadronic mass matrix.

To derive the results of the quark model for baryon states,
we shall make use of two lemmas, that may also have
applications in other contexts.\\

\noindent
{\sc Lemma 1}: Any set of hadronic states that furnish a
representation of the commutation relations (17) and (18),
in which for each helicity  any given isospin appears at
most once, must be degenerate.\\
\noindent
Proof:  Eq. (18) may be written as
$[X_a(\lambda),[X_b(\lambda),m^2]=m_4^2(\lambda)\delta_{ab}
$.
By taking the commutator of this with $X_c(\lambda)$ and
using the Jacobi identity, it is easy to see that
$[X_c(\lambda),m_4^2(\lambda)]=[X_c(\lambda),m^2].
$
Hence the mass-squared matrix may be written
$m^2=m_4^2(\lambda)+m_0^2(\lambda)
$, where $m_0^2(\lambda)$ is a chiral scalar, satisfying
$
[X_c(\lambda),m_0^2]=0
$.  Also, $m_4^2(\lambda)$ and $m_b^2(\lambda)\equiv
[X_b(\lambda),m_4^2(\lambda)]$ form a chiral four-vector, in
the sense that
$[X_a(\lambda), m_b^2(\lambda)]=m_4^2(\lambda)\delta_{ab}$.
Now, since we assume that for a given helicity, each isospin
occurs just once, each isospin for a given helicity can come
from just one irreducible
representation of $SU(2) \times SU(2)$.  The mass $m$
commutes with isospin, so  $m^2$ can have matrix elements
only between baryonic
states with the same isospin, and hence belonging to the
same
representation of $SU(2) \times SU(2)$.  But a
$(1/2,1/2)$ operator like $m_4^2(\lambda)$ can have no
matrix elements between
two states that belong to the same irreducible
representation
$(A,B)$ of $SU(2) \times SU(2) $, so in this representation
all matrix elements of
$m^2_4$ must vanish.  This leaves us with
$m^2=m_0^2$, and since this commutes with $X_a$ all hadron
states connected by one-pion transitions must have the same
mass.\\

{\sc Lemma 2}: Any set of degenerate hadronic states of the
same parity that furnish a representation of the commutation
relations (17) and (19) also furnish a representation of an
$SU(4)\times O(3)$ algebra with $SU(4)$ generators $T_a$,
$S_\alpha$, and $D_{ai}$ and $O(3)$ generators
$\tilde{S}_i\equiv J_i-S_i$, satisfying the commutation
relations\footnote{We use $a,\; b,\; c,$  etc. for isovector
indices and $i,\;j,\;k$, etc. for spatial vector indices.}
\begin{eqnarray}
&&[T_a,T_b]=i\epsilon_{abc}T_c~~~~~~[S_i,S_j]=i\epsilon_{ijk
}S_k~~~~~~[T_a,S_i]=0 \\&&
[T_a,D_{bi}]=i\epsilon_{abc}D_{ci}~~~~~~~[S_i,D_{aj}]=i
\epsilon_{ijk}D_{ak} \\&&
[D_{ai},D_{bj}]=i\delta_{ij}\epsilon_{abc}T_c+i\delta_{ab}
\epsilon_{ijk}S_k\\&&
[\tilde{S}_i,\tilde{S}_j]=i\epsilon_{ijk}\tilde{S}_k\\&&
[\tilde{S}_i,T_a]=[\tilde{S}_i,S_j]=[\tilde{S}_i,D_{aj}]=0
\end{eqnarray}
where
\begin{equation}
[D_{a3}]_{\lambda'\beta,\lambda
\alpha}=\delta_{\lambda'\lambda }
[X_a(\lambda)]_{\beta\alpha}
\end{equation}
and $J_i=\tilde{S}_i+S_i$ is the usual spin matrix acting on
helicity indices, with
\begin{equation}
[J_{3}]_{\lambda'\beta,\lambda
\alpha}=\delta_{\lambda'\lambda }\delta_{\beta\alpha}
\lambda\;.
\end{equation}
Proof:  For a transition between equal parity states, in the
rest frame of the initial particle the invariant pion
transition amplitude $<\beta,{\bf p}^\prime, \lambda^\prime
;  \pi_a,q|{  }S{ }
|\alpha,{\bf p},\lambda>\times\sqrt{q^0p'^0p^0}$ must be an
odd function of the momentum of the final particle, whose
magnitude  is proportional to $m_\alpha^2-m_\beta^2$.  Hence
for $m_\alpha^2-m_\beta^2\rightarrow 0$, this invariant
amplitude must be proportional to a linear combination of
components of the final momentum vector.  Rotational
invariance requires that the coefficients must also form a
three-vector.  In particular, for ${\bf p}'$ in the 3-
direction the coefficient of $p'_3$, which is proportional
to $X_a$,  must be the third component of a quantity that
transforms like a spatial 3-vector and an isovector, in the
sense that $\delta_{\lambda',\lambda }
[X_a(\lambda)]_{\beta\alpha}=[D_{a3}]_{\lambda'\beta,\lambda
\alpha}$ where $[T_a,D_{bi}]=i\epsilon_{abc}D_{ci}$ and
$[J_i,D_{aj}]=i
\epsilon_{ijk}D_{ak}$.
Next, we must consider the commutation relations of the
$D_{ai}$ with each other.
The commutators of two
$D$'s may be written as a sum of two terms, one symmetric in
space indices and antisymmetric in isospin indices, and the
other vice versa:
$$
[D_{ai},D_{bj}]=i\epsilon_{abc}A_{ij,c}+i\epsilon_{ijk}
B_{ab,k}
$$
with $A_{ij,c}=A_{ji,c}$ and $B_{ab,k}=B_{ba,k}$.  The
commutation relation (17) now takes the form $A_{33,a}=T_a$.
{}From rotational invariance (or formally, by taking repeated
commutators with $J_i$), we easily see then that $
A_{ij,a}=\delta_{ij}T_a$.  To find $B_{ab,k}$, we must use
the spin-flip
superconvergence relation (19), which can be rewritten in
the form $ [D_{a3},[D_{b3},m(J_1\pm iJ_2)]]\propto
\delta_{ab}$,
or, since the states connected by $D_{ai}$ are degenerate, $
[D_{a3},D_{b1}\pm i D_{b2}]\propto \delta_{ab}$.
It follows that $B_{ab,2}$ and $B_{ab,1}$ are proportional
to $\delta_{ab}$, and by rotational invariance the same is
true of $B_{ab,3}$, so\footnote{This result was obtained in
reference [19] by a weak argument, that the algebra
containing the pion transition amplitudes should not contain
any $T=2$ operators.  The present argument, based on the
superconvergence relation (19), avoids this hand-waving.}
$B_{ab,i}=\delta_{ab}
S_i$, verifying Eq. (22).
{}From (22), we have
$S_i=-i\epsilon_{ijk}[D_{aj},D_{ak}]/6$.
Using (22) again to calculate the commutator of this with
$D_{bj}$, we easily obtain the commutator
$[S_i,D_{bj}]=i\epsilon_{ijk} D_{bk}$, and using this
together with the above expression for $S_j$ in terms of the
$D$'s, we also find
$[S_i,S_j]=\epsilon_{ijk}S_k$, verifying the remainder of
the commutators (20)-(21).
We have already mentioned that $D_{aj}$ is a 3-vector, in
the sense that $[J_i,D_{aj}]=i\epsilon_{ijk}D_{ak}$, so the
same is true of $S_j$. It follows then that
$\tilde{S}_i\equiv J_i-S_i$ satisfies the commutation
relations (23) and (24).\\

To apply Lemma 1 to the tower states, we note that for each
helicity $\lambda$, the tower
contains isospins $T=|\lambda|,\, |\lambda|+1,\cdots N_c/2$.
Since each isospin occurs just once, for a given helicity
each isospin can come from just one irreducible
representation of $SU(2) \times SU(2)$.  Therefore according
to Lemma 1, the tower states must be degenerate.

Lemma 2 then tells us that the pion transition amplitudes
are part of an $SU(4)\times O(3)$ algebra.
It is easy to see  that the baryons transform under
$SU(4)\times O(3)$
as a symmetric $SU(4)$ tensor\footnote{There are actually
two of these
representations, the contravariant and covariant tensors of
rank $N_c$.  They differ only in the sign of $D_{ai}$, so we
can choose either of these representations by adjusting the
sign of the one-pion state.} of rank $N_c$ and an $SO(3)$
singlet, because this is the only
representation of $SU(4)\times O(3)$ that contains just the
spins and isospin states of the baryon tower.     The
matrix elements of
$D_{ai}$ in this representation may be calculated by
representing it by
$\sum \sigma_i t_a$, just as in the non-relativistic quark
model.
{\em Thus once we assume
that general Adler-Weisberger and superconvergence sum rules
are saturated by the tower states,  the other
consequences of the non-relativistic quark model for pion
transitions and $g_A$ follow
immediately from these sum rules, with no further need for
the large $N_c$ approximation.}

All of the above results apply also for baryons that contain
some number $N_h$ of heavy quarks.  Since pion transition
amplitudes and baryon masses are independent of the spin
$3$-component of the heavy quarks, it is only necessary to
replace $N_c$ everywhere above with $N_c-N_h$.

I am grateful for helpful conversations about the large
$N_c$ approximation with Howard Georgi, Vadim Kaplunovsky,
and Aneesh Manohar.

\pagebreak
\noindent
{\bf References}

\begin{enumerate}

\item The terms in (1) involving pions alone or pions and a
single nucleon bilinear were given by S. Weinberg, Phys.
Rev. Lett. {\bf 18}, 1 88 (1967).  It was realized later
that the terms involving $C_S$ and $C_T$ make contributions
of the same order in small energies: S. Weinberg, Phys.
Lett. {\bf B251},
288 (1990); Nucl. Phys. {\bf B363}, 3 (1991).

\item S. Weinberg, Physica {\bf 96A}, 327 (1979).

\item J. Gasser and H. Leutwyler, Phys. Lett. {\bf 125B},
321, 325 (1985); Ann. Phys. {\bf 158}, 142 (1984).

\item V. Bernard, N. Kaiser, and U-G. Meissner, Nucl. Phys.
{\bf B 383}, 442 (1992); Ulf-G. Meissner, Lectures delivered
at the XXXII. Internationale Universit\"{a}tswochen f\"{u}r
Kern- und Teilchenphsik, Schladming, February 24 - March 6,
1993, hep-ph 9303298.

\item C. Ord\'{o}\~{n}ez and U. van Kolck, Phys. Lett. B
{\bf 291}, 459 (1992); C. Ord\'{o}\~{n}ez, L. Ray, and  and
U. van Kolck, Phys. Rev. Lett. {\bf 72}, 1982 (1994).

\item S. Weinberg, Transactions of the N. Y. Academy of
Sciences {\bf 38}, 185 (1977).

\item After this talk was presented, H. Leutwyler informed
me of a paper by A. Krause, Helv. Phys. Acta {\bf 63}, 3
(1990).  This paper  gave an $SU(3)\times SU(3)$ effective
Lagrangian that
includes a mass term from which it would be possible to
derive the $B$ term in (4), but
did not consider the implications of this Lagrangian for
pion nucleon interactions.

\item S. Weinberg, Phys. Rev.  {\bf 166}, 1568 (1968).

\item S. Coleman, J. Wess and B. Zumino, Phys. Rev. {\bf
177},
2239 (1969) ; C.G. Callan, S. Coleman, J. Wess and B.
Zumino,
Phys. Rev. {\bf 177}, 2247 (1969).

\item J. Wess and B. Zumino, Phys. Lett.  {\bf 37B},  95
(1971).

\item E. Witten,  Nucl. Phys. {\bf B223}, 422 (1983).

\item E. D'Hoker and S. Weinberg, UCLA-Texas preprint, to be
published in Physical Review D.

\item G. 't Hooft, Nucl. Phys. {\bf B 72}, 461 (1974); S.
Coleman, in {\em Aspects of Symmetry} (Cambridge University
Press, Cambridge, 1985).

\item E. Witten, Nucl. Phys. {\bf B160}, 57 (1979).

\item J.-L. Gervais and B. Sakita, Phys. Rev. Lett. {\bf
52}, 87 (1984).

\item R. Dashen and A. V. Manohar, Phys. Lett. {\bf B315},
425, 438 (1993); E. Jenkins, Phys. Lett. {\bf 315}, 431, 447
(1993); R. Dashen, E. Jenkins, and A. V. Manohar, Phys. Rev.
D {\bf 49}, 4713 (1994).

\item C. D. Carone, H. Georgi, and S. Osofsky, Phys. Lett.
{\bf F 322}, 227 (1994); C. D. Carone, H. Georgi, L. Kaplan,
and D. Morin, Harvard preprint HUTP-94/A008 (1994).

\item A. Wirzba, M. Kirchbach, and D. O. Riska, Darmstadt-
Helsinki preprint (1993), hep-ph/9311299.

\item S. Weinberg, Phys. Rev. {\bf 177},
2604 (1969).  Some algebraic errors in the Appendix to this
paper are corrected here.

\item S. Weinberg, Phys. Rev. Lett. {\bf 22}, 1023 (1969).

\item S. Weinberg, in {\it Lectures on Elementary
Particles and Quantum Field Theory, 1970 Brandeis University
Summer Institute on Theoretical Physics}
(M.I.T. Press, Cambridge, MA, 1970), pp. 285-393.

\end{enumerate}
\end{document}